\documentclass[prd,showpacs,preprintnumbers,twocolumn,superscriptaddress,floatfix]{revtex4-1}
\usepackage[utf8]{inputenc}
\usepackage{bm,amsmath,amssymb}
\usepackage{graphicx}
\usepackage{subfigure}
\usepackage{color}
\usepackage{hyperref}
\usepackage{multirow}
\usepackage[dvipsnames]{xcolor}
\usepackage{soul}
\usepackage{comment}
\usepackage[normalem]{ulem}
\let\oldalign\align
\def\align{\linenomath\oldalign}
\begin{document}
%\linenumbers

\title {Imprints of $U_A(1)$ chiral anomaly and disorder in the Dirac eigenspectrum of QCD at finite temperature  }

\author{Ravi Shanker}
\email{Corresponding author: rshanker@imsc.res.in}
\affiliation{The Institute of Mathematical Sciences, Chennai 600113, India}
\affiliation{Homi Bhabha National Institute, Training School Complex, Anushaktinagar, Mumbai 400094, India}
\author{Harshit Pandey}
\affiliation{The Institute of Mathematical Sciences, Chennai 600113, India}
\affiliation{Homi Bhabha National Institute, Training School Complex, Anushaktinagar, Mumbai 400094, India}
\author{Sayantan Sharma}
%\email{sayantans@imsc.res.in}
\affiliation{The Institute of Mathematical Sciences, Chennai 600113, India}
\affiliation{Homi Bhabha National Institute, Training School Complex, Anushaktinagar, Mumbai 400094, India}

%\date{\today}
%------------------------------------------------------------------------------------
%       abstract
%------------------------------------------------------------------------------------
\begin{abstract}

We perform a comprehensive study of the properties of the Dirac eigenvalue spectrum in QCD as a function of  
temperature on the lattice. In addition to effects due to the interplay between interactions and disorder 
inherently present in a many-body system, the Dirac spectrum also contains crucial information about the 
effective restoration of different subgroups of almost exact two-flavor chiral symmetry in QCD. We calculate 
the infrared eigenvalues of the overlap Dirac operator on 2+1 flavor QCD ensembles generated using domain wall 
fermion discretization, on a large volume lattice. From the normalized level spacing ratios, we identify those 
eigenvalues that have intermediate level statistics, distinctly different from the majority in the 
bulk spectrum that follow universal level fluctuations similar to a random matrix of Gaussian unitary type. We 
provide an explanation of these intermediate level ratios in terms of a specific random matrix model and 
quantify the correlation between these eigenstates and disorder in the gauge fields manifested in the renormalized 
Polyakov loop values. Whereas the existence of \emph{intermediate} eigenmodes is intimately connected to 
the \emph{effective} restoration of different subgroups of chiral symmetry close to the chiral crossover transition, their  
origin can be traced to random uncorrelated disorder at higher temperatures when the $U_A(1)$ is 
\emph{effectively restored}. We also, for the first time, calculate the Thouless conductance for the Dirac 
spectrum that quantifies the structural rigidity of the eigenvectors, and use it as a diagnostic tool to 
understand the restoration of the anomalous $U_A(1)$ subgroup of chiral symmetry and localization driven by 
disorder.

%Only when the U(1) is effectively restored can one identify the effects of localization in the Dirac spectrum of QCD.

\end{abstract}

\pacs{  12.38.Gc, 11.15.Ha, 11.30.Rd, 11.15.Kc}
\maketitle

%-----------------------------------------------------------------------------------------------------
%-----------------------------------------------------------------------------------------------------

\section{Introduction}

Interacting many-body quantum systems continue to challenge our understanding of fundamental 
processes like thermalization and transport. Understanding the universal properties of the eigenvalue 
spectra of the Hamiltonians describing such systems provides valuable insights into their 
dynamics. The universal fluctuations of the level spacings between adjacent energy eigenvalues 
are one such observable. Bohigas-Giannoni-Schmit conjecture~\cite{Bohigas:1983er} states 
that for quantum systems whose classical dynamics is non-integrable, the universal fluctuations of 
the level separations between eigenvalues of the Hamiltonian belong to one of the three universality 
classes of Wigner-Dyson statistics within Random Matrix theory (RMT). On the other hand, if the corresponding
classical dynamics is completely integrable, then the quantum energy eigenvalues behave like a sequence of 
independent random variables according to the Berry-Tabor conjecture~\cite{10.1098/rspa.1977.0140}. 
Therefore, studying the universal level spacing fluctuations of a quantum system allows us to learn 
about the classical dynamics and vice versa; in particular, it enables us to understand how classical 
(non-)integrability manifests itself in the quantum many-body theory. This is important since it is 
expected that the long-time dynamics in classical non-integrable many-body systems, characterized by 
ergodicity and mixing in the phase space, is universally governed by a thermal fixed point. However, the 
connection between chaos and thermalization is not very apparent in a quantum system~\cite{Deutsch:2018ulr}.

Many realistic physical systems far from integrability exhibit quasi-periodic motion which is challenging 
to understand even from a purely classical point of view, let alone understanding the classical-quantum 
correspondence in these systems~\cite{Bohigas:1992hj}. Some ideas inspired by semiclassical correspondence 
have been put forward by Berry and Robnik, who proposed that the spectral statistics of such mixed systems 
should be understood as arising from an uncorrelated superposition of spectra from different universality 
classes, each spectrum being associated with chaotic or regular regions in the classical phase 
space~\cite{MVBerry_1984}. However, extensive investigations~\cite{Bohigas:1992hj,TProsen_1994,PhysRevE.97.062212} 
reveal that the universal fluctuations of the energy spectrum, characterized in terms of integrable (ordered) 
as well as non-integrable (disordered) regions might be correlated, even in cases where no classical 
analog might exist~\cite{Nandkishore:2014kca,Serbyn:2015lmq,Amini:2015mvz,Moudgalya:2021ixk,Moudgalya:2021xlu,Adler:2024gmz} or the ordered (integrable) spectra may become chaotic due to strong admixture.  
There can be other non-trivial scenarios as well. Many-body systems can undergo localization transitions 
when symmetries of the Hamiltonian are broken as a function of external parameters, e.g., the strength 
of disorder, leading to a change in the nature of level fluctuations. Such cases are typically addressed by 
random matrix ensembles which interpolate between Poisson and Wigner-Dyson level statistics as observed 
in localization transitions, e.g., close to the mobility edge~\cite{PhysRevB.47.11487} in the Anderson 
model~\cite{Anderson:1958vr}.  In this particular case, the critical eigenstates for very closely spaced 
eigenvalues are strongly correlated and have significant overlap with each other despite their sparse 
fractal-like structure. However, very small correlations exist between eigenvalues with large separations.

A prominent example of a strongly interacting many-body quantum system, which has all these features, is 
described by a non-Abelian gauge theory known as Quantum Chromodynamics (QCD). The QCD Hamiltonian is 
difficult to realize in three dimensions (see e.g., Ref.~\cite{Das:2025utp} and references therein); 
instead, we study the spectral properties of the QCD Dirac operator in Euclidean 3+1 dimensions.  The Dirac 
eigenspectrum consisting of eigenstates with different amounts of localization not only determines the transport 
properties in QCD but also explains key features of a non-trivial thermodynamic chiral crossover 
transition~\cite{Sharma:2018y2, Lombardo:2020bvn} which is an exact phase transition in the chiral limit. 
Thus, a strongly interacting many-body system described by QCD is an important example of a theorist's laboratory 
to study how, in addition to the interplay between disorder and interactions, the correlations due to the criticality 
related to the chiral phase transition get imprinted in the level spacing statistics of the Dirac eigenvalues. This 
present work is a detailed study of the QCD Dirac spectrum which provides new insights and ways to disentangle the 
features due to changes in the localization (transport) properties as a function of temperature from those due to 
spontaneous breaking of different subgroups of chiral symmetry.

\section{Background}

The eigenvalue spectrum of the Dirac operator provides important insights regarding the symmetries of 
QCD and their breaking. An eigenstate of the Dirac operator describes a one-particle massless fermion 
(quark) state in the presence of a gauge background.  The gauge fields provide both the interactions and the disorder 
whose strength varies as a function of temperature.  Properties of the eigenvalue spectrum of the QCD Dirac 
operator, in presence of gauge ensembles, have been studied extensively using lattice QCD at zero as well 
as finite temperatures.  The density of vanishingly small eigenvalues of the Dirac operator in the 
chiral symmetry broken phase gives an accurate estimate of the chiral condensate, the order parameter of the 
chiral phase transition in the limit of massless quarks~\cite{Smilga:1993in}. At temperatures $T<T_{pc}$, where 
$T_{pc}\sim 156.5$ MeV is the pseudo-critical temperature~\cite{HotQCD:2018pds,Burger:2018fvb,Borsanyi:2020fev} 
corresponding to the chiral crossover transition in QCD with physical quark flavors, 
the level statistics of the deep infrared part of the spectrum can be explained within the universality class 
of RMT consisting of chiral Gaussian unitary ensemble (GUE)~\cite{Verbaarschot:1994qf,Verbaarschot:1993pm, Verbaarschot:1996qm,Kanazawa:2018kbo} whereas for the higher eigenvalues the universal 
features can be explained within Wigner's classification with a RMT ensemble of conventional Gaussian 
unitary type. 

At higher temperatures, the non-singlet part of the almost exact two-flavor chiral symmetry is 
\emph{effectively} restored, i.e. the disconnected part of the chiral susceptibility matches with 
the topological susceptibility up to $\mathcal{O}(m^2)$, where $m$ is the mass of the light quark 
flavor~\cite{Petreczky:2016vrs, Gavai:2024mcj}. This also leads to perceptible changes in the eigenvalue 
spectrum of the Dirac operator--the intercept vanishes, leading to the formation of a small peak-like feature 
in the infrared part of the eigenvalue spectrum. This peak has been observed in studies using a mixed-action 
set-up~\cite{Dick:2015twa,Sharma:2016cmz,Holicki:2018sms} which was debated for a 
while~\cite{Tomiya:2016jwr, Aoki:2020noz, Aoki:2021qws, Chiu:2014Ld}. However, existence of the peak-like 
feature was shown with dynamical domain wall fermions~\cite{Buchoff:2013nra} on a finite lattice, now also 
observed to survive in the continuum limit in 2+1 QCD with HISQ discretization~\cite{Kaczmarek:2023bxb}. 
The higher (bulk) eigenvalues denoted by $\lambda$, whose density grows linearly with $\lambda$, are disordered 
i.e., their level spacing ratios are consistent with the predictions from RMT belonging to the GUE. 
The peak-like feature and bulk spectrum overlap strongly, resulting in the most infrared
eigenmodes to be delocalized~\cite{Alexandru:2021pap,Alexandru:2023xho,Alexandru:2024tel} 
but not ergodic, since their inverse participation ratio does not scale inversely with volume. 
Their level spacing fluctuations are intermediate between the Poissonian 
and GUE values~\cite{Pandey:2024goi}. These infrared eigenmodes are characteristic 
of a strongly-coupled regime~\cite{Alexandru:2015fxa,Alexandru:2019gdm,Alexandru:2021pap} 
and exhibit fractal-like features~\cite{Pandey:2024goi}. The median of the distribution of 
fractal dimensions for these \emph{intermediate} eigenvalues is tantalizingly similar to 
the expectations from an O(4) spin model~\cite{Pandey:2024goi}. In contrast, bulk 
eigenstates are ergodic since their inverse participation ratios scale as 
$\sim L^{-2.8}$~\cite{Pandey:2024goi} and are thus delocalized over the entire volume.

The $U_A(1)$ subgroup will be \emph{effectively} restored at a 
temperature, $T_{U(1)}$, when physical observables which are sensitive to it, e.g., the topological 
susceptibility has a temperature dependence which is consistent with the expectation from a dilute 
gas of instantons. This can be motivated by the fact that once an onset to the dilute instanton gas regime 
occurs, there are no further changes in the topological properties even when temperatures are 
increased to asymptotically large values. The $T_{U(1)}$ is close to $\gtrsim 1.5~T_{pc}$ inferred from the 
temperature dependence of the topological susceptibility in QCD~\cite{Petreczky:2016vrs}, 
from the features in the Dirac eigenvalue spectrum~\cite{Dick:2015twa,Alexandru:2024tel}, from 
the degeneracy of temporal meson correlators~\cite{Bignell:2026ybw} and through tensor singlet 
channels~\cite{Chiu:2026sxy}. A weaker bound $T_{U(1)}\gtrsim 1.2~T_{pc}$ was obtained from the 
fact that the value of $\chi_\pi-\chi_\delta$, arising due to the mixing between near-zero 
eigenvalues with the bulk, vanishes at this temperature~\cite{Kaczmarek:2023bxb}.

The scenario at even higher temperatures $T>T_{U(1)}$ is not yet well understood. Whereas a peak-like 
feature of near-zero eigenmodes similar to $m^2\delta(\lambda)$ should survive, expected from a dilute gas-like 
scenario for instantons~\cite{Dick:2015twa,Ding:2020xlj,Kovacs:2023vzi,Alexandru:2024tel}, it is heavily suppressed 
as a function of temperature. Thus its contribution to topological susceptibility decreases as $\sim T^{-8}$
~\cite{Petreczky:2016vrs,Borsanyi:2016ksw,Bonati:2018blm}. Eventually, these infrared modes should be gapped 
out from the bulk part of the eigenvalue spectrum. Recent studies indicate that at sufficiently high temperatures, 
the most infrared part of this bulk spectrum contains localized eigenvalues which can be distinguished 
from the other delocalized bulk eigenmodes, which follow GUE level statistics, through a mobility 
edge~\cite{Holicki:2018sms,Giordano:2016nuu,Kehr:2023wrs}. This phenomenon is analogous to a localization 
transition~\cite{Garcia-Garcia:2006vlk,Kovacs:2010wx,Giordano:2013taa,Ujfalusi:2015nha,Cossu:2016scb,Holicki:2018sms, 
Kehr:2023wrs} observed in the Anderson model~\cite{Anderson:1958vr}. Nonetheless, a naive Anderson-like scaling 
of its temperature dependence demonstrates that the mobility edge might vanish at the pseudo-critical 
temperature~\cite{Holicki:2018sms,Garcia-Garcia:2006vlk,Giordano:2026bqj}, but it might also disappear only 
at the critical temperature which is achieved in the chiral limit~\cite{Kehr:2023wrs}. The robustness of these 
observations will in the future be tested through carefully performed infinite volume and continuum extrapolations.

To summarize, whereas there is a wealth of results on spectral signatures related to  
\emph{effective} restoration of the different subgroups of chiral symmetry, and on the 
localization properties akin to an Anderson-like transition, separately, the interplay between 
these two physically distinct phenomena is not so clear. We want to emphasize here that these 
two phenomena have distinct physical origins and thus their consequences will also be different. 
Anderson transition is a quantum phase transition which, strictly speaking, is a zero-temperature 
phenomenon. A possible way to understand such a (de)-localization transition at finite temperatures 
is to attribute the change in the strength of the disorder due to gauge fluctuations arising due 
to the change in temperature; however, a clear connection between the two is not very apparent. 
In QCD, varying the temperature also results in the sequential (effective) restoration of different 
subgroups of chiral symmetry. Evidence suggests that the restoration of the $SU_A(2)$ occurs through 
a second-order phase transition in the chiral limit~\cite{HotQCD:2019xnw}, remnant effects of which will 
remain manifest in the Dirac eigenvalue spectrum even for physical light quarks~\cite{Pandey:2024goi}. 
The effective restoration of $U_A(1)$ will result in characteristic changes in the localization properties 
of the Dirac eigenvectors; on the other hand, an Anderson-like transition also results in structural changes 
in the eigenvectors of Dirac operator. It is thus imperative to disentangle these two phenomena, to unambiguously 
identify the mechanism that drives these effects. However, the temperature regimes where these occur overlap with 
each other, which makes it challenging to unambiguously separate these two effects.

\section{An outline of this work}

In this work, we provide a systematic way of disentangling the impact of \emph{effective} restoration 
of $U_A(1)$ from the gradual onset of localization driven by disorder in 1-particle quark states through 
a detailed study of the QCD Dirac eigenspectrum on the lattice. Though both these distinct physical 
phenomena result in the presence of infrared eigenvalues with intermediate level ratios at temperatures 
above $T_{pc}$, we discuss discerning observables that can distinguish between the two. We demonstrate that the 
onset of localization at high temperatures due to disorder is distinct from the localization largely driven 
by topological fluctuations just above $T_{pc}$ and can only be observed when the $U_A(1)$ subgroup of 
chiral symmetry is \emph{effectively} restored.

We also measure for the first time the structural rigidity of the Dirac eigenstates 
in terms of their curvature arising as a result of applying a twist on one of the spatial 
boundaries of the lattice box. It is well known that the curvature of eigenvalues of 
the Hamiltonian is related to the Thouless conductance~\cite{JTEdwards_1972}. 
We propose that the analogue of Thouless conductance for the Dirac eigenspectrum can 
be used as a diagnostic tool for characterizing the \emph{effective} restoration of 
$U_A(1)$.

\section{Details about the numerical techniques}
\label{sec:Num}

The gauge configurations for QCD with two light and a strange quark flavor were 
generated using M\"{o}bius domain wall discretization for fermions and Iwasaki gauge action. 
The computations were performed on a spacetime lattice which has $N=32$ sites 
along each of the three spatial directions and $N_\tau=8$ sites along the Euclidean 
temporal direction. We focus on the high temperature phase where the non-singlet part of 
the (almost) exact two-flavor chiral symmetry in QCD is \emph{effectively} restored, whereas its 
$U_A(1)$ subgroup remains broken but the magnitude of this breaking has a characteristic 
dependence on the temperature. At temperatures $T=164$-$339$ MeV, we study the spectral properties 
of the QCD Dirac operator, realized on the lattice with the overlap Dirac operator.  
The gauge configurations for $T \lesssim 181$ MeV were already generated and used for calculating 
the disconnected piece of the chiral susceptibility and the pseudo-critical temperature 
$T_{pc}\simeq 158.7$ MeV~\cite{Gavai:2024mcj}. The temperatures $T=1/(N_\tau a)$ in physical units 
are expressed in terms of the Sommer parameter $r_0$ at each value of the gauge coupling $g$ by
relating it to the lattice spacing $a$ through the two-loop beta-function~\cite{Lin:2014tym}.
However all gauge ensembles characterized by the bare parameter $\beta=6/g^2$ for $\beta \gtrsim 1.829 $ 
were generated from scratch. The temperature corresponding to $\beta=1.829$ was already estimated 
in Ref.~\cite{Lin:2014tym}. For precisely determining the temperature in MeV units for the  
larger $\beta$ values, we have first extracted the lattice spacing $a$ in units of the mass of 
the omega baryon $m_{\Omega}$~\cite{Lin:2014tym,RBC:2014ntl,Jung:2022sjp} for a wide range of 
$\beta=1.70$-$ 2.31$. We next performed a fit of these extracted inverse spacings as a 
function of $\beta$ and then used this fit and the 2-loop beta function to extract the corresponding 
temperatures in MeV for all $\beta$ values. We checked that the temperature we extracted from this 
fit, corresponding to a lower $\beta=1.633$, matches the value reported earlier~\cite{Bhattacharya:2014ara}. 
Details of the parameters used in the configuration generation, e.g. the M\"{o}bius parameter $c_5$, the 
extent of the fifth dimension $L_5$ and the number of eigenvalues computed for each statistically independent 
configuration at different temperatures are mentioned in Table.~\ref{tab:table1}. The quark masses are physical 
which are fixed by setting the value of the pion mass to $140$ MeV and the kaon mass to $435$ 
MeV~\cite{Gavai:2024mcj}.  The spatial lattice size in physical units varies between $4$-$2.3$ fm 
for $T=195$-$339$ MeV. 
In order to check the volume dependence of our results, we also generated gauge configurations 
on a $24^3\times 8$ lattice at $T=195$ MeV and a $40^3\times 8$ lattice at $T=314$ MeV, also listed 
in Table \ref{tab:table1}. This was to check the finite volume effects for one temperature which is 
higher than $T_{U(1)}$ and one below it. 

As mentioned in Table~\ref{tab:table1}, we typically calculate the first $100$-$200$ non-zero eigenvalues 
of $D_{ov}D^\dagger_{ov}$, where $D_{ov}$ is the massless overlap Dirac operator on each of these gauge 
configurations.  This choice of the lattice Dirac operator is motivated by the fact that the 
overlap fermions have an exact chiral symmetry on the lattice, satisfy an index theorem and do not have 
additional lattice artifacts due to the breaking of chiral symmetry, as in the case of Wilson fermions or the 
breaking of the continuum spin and flavor symmetries in staggered fermions. We have numerically implemented 
the overlap Dirac operator by representing the matrix sign function exactly in terms of the first $30$ 
eigenvalues of $D_w^\dagger D_w$ and approximating it by a Zolotarev rational polynomial with $25$ terms 
in the rest of the vector space. 
The numerical accuracy of our procedure ensures that the resultant overlap Dirac operator satisfies the 
Ginsparg-Wilson relation to a precision of $\lesssim 10^{-9}$ on each gauge configuration we have studied 
and the sign function is implemented to a numerical precision of $10^{-10}$. The eigenvalues of the overlap 
Dirac operator are calculated numerically using the Kalkreuter-Simma Ritz algorithm~\cite{Kalkreuter:1995mm} 
with restarts. For optimization, we have first estimated the number and chirality of zero modes present per 
configuration and used this information to project the $D_{ov}D^\dagger_{ov}$ onto the opposite chirality sector. 
We then calculate the first $100$-$200$ eigenvalues of this projected overlap Dirac operator and extract the 
eigenvalues of $D_{ov}$ from them. We henceforth denote the imaginary part of the $n$-th eigenvalue as $\lambda_n$. 
We have also checked that the real part of the first $\mathcal{O}(200)$ overlap eigenvalues is about two orders of 
magnitude smaller and hence sub-dominant compared to their imaginary part. This allows us to directly relate our 
study to the continuum case where the Dirac operator has only imaginary eigenvalues.

\begin{table}[ht]
 \centering
 \begin{tabular}{|c|r|r|r|r|r|c|c|c|}
 \hline
 \hline
$T$ (MeV) & $\beta ~~$ & $c_5$& $L_5$&$N_s$ & $N_\tau$ & L (fm)  & $N_{\text{eigen}}$ & $N_\text{confs}$ \\
 \hline
 \hline
  164 & 1.725 & 1.5 & 16 &32 & 8 & 4.81 &100 & 95 \\
   \hline
  177 & 1.771 &1.0 & 16&32 & 8 & 4.45 &100 & 93 \\
   \hline
  195 & 1.829 &0.9 & 16&32 & 8 & 4.04 &100 &108 \\
  
   &  & &  & 24 & 8 & 3.03 & &177 \\
   \hline
  270 & 2.130 &0.5 &24&32 & 8 & 2.92  &150 &122 \\
   \hline
  314 & 2.250 &0.5 &12&32 & 8 & 2.51 &200 &111 \\
      &  & & & 40 & 8 & 3.14 & &27 \\
   \hline
  339 & 2.310 &0.5 &12&32 & 8 & 2.32 &200 &92 \\
\hline
  \end{tabular}
  \caption{Number of gauge configurations at different temperatures, \( \beta =6/g^2\), lattice size $N_s, N_\tau$ and the spatial length $L=N_s a$ in fm, used in this work. The table also lists the number of overlap Dirac eigenvalues calculated per gauge configuration, along with the values of the M\"{o}bius parameter \( c_5 \)  and \( L_5 \) used in the generation of gauge ensembles at each temperature. }
  \label{tab:table1}
\end{table}

We have also calculated the correlation between the real part of the renormalized Polyakov loop values and these first 
$\mathcal{O}(100)$ Dirac eigenstates.  We performed our calculation using the gradient flow renormalization scheme 
using the Wilson gauge action~\cite{Luscher:2010iy}. The flow equation is solved using the fourth-order Runge-Kutta 
scheme with a step size of $0.01$  up to a flow time $t$ which should ideally be determined from the criterion
$a\ll \sqrt{8t} \ll 1/T $. In order to ensure that this criterion is satisfied for all temperatures we have 
studied, the flow time in physical units was chosen to be $\sqrt{8t} = 0.44$ fm for $T \leq 195$ MeV and 
$\sqrt{8t} = 0.25$ fm for higher temperatures $T > 200$ MeV. We applied a constant shift in free energy to 
match with the earlier continuum extrapolated results from Ref.~\cite{Bazavov:2016uvm} at the highest temperature 
in order to fix the scale. This same shift in the free energy values for two lower 
temperatures $270$ and $314$ MeV also gave us an estimate of the renormalized Polyakov loop, which is in good 
agreement with the continuum results mentioned in Ref.~\cite{Bazavov:2016uvm}.

\section{Results}

\subsection{Level spacing ratios near the chiral crossover transition }

\begin{figure}[h]
    \centering
    \raisebox{-\height}{\includegraphics[width=0.48\textwidth]{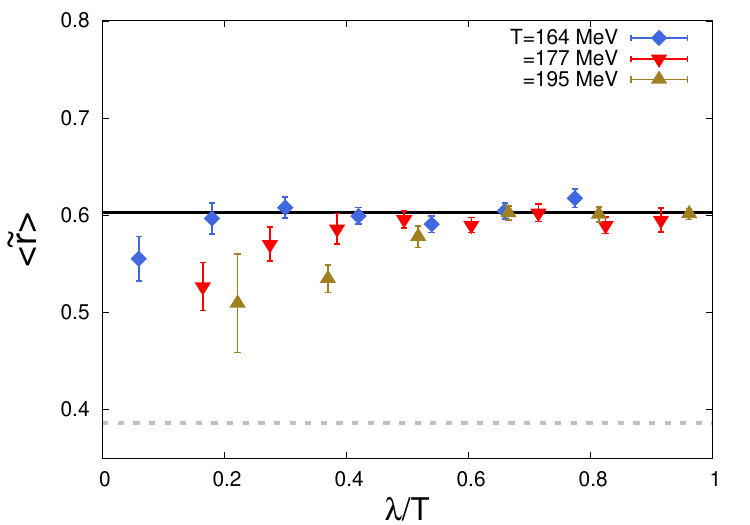}}
    \caption{The variation of $\langle \tilde{r} \rangle$ in bins of the Dirac eigenvalues $\lambda/T$ in 
    the temperature range where the non-singlet part of chiral symmetry is \emph{effectively} restored but not its $U_A(1)$ subgroup. The solid and dashed lines represent the expected $\langle \tilde{r} \rangle$ for the 
    eigenvalues with GUE and Poisson level statistics, respectively.}
    \label{fig:rtildavslambda}
\end{figure}

We first calculate the spacings between the neighboring eigenvalues of the Dirac operator, 
denoted as $s_n=\lambda_{n+1}-\lambda_n$. Instead of calculating the level 
spacing distribution, we study the normalized ratios of the consecutive level spacings defined as 
$\tilde{r}_n=\text{min}(r_n,1/r_n)$, where $r_n = s_{n+1}/s_n$. Such a construction eliminates the 
local dependence of spacing distributions on the eigenvalue level density~\cite{PhysRevB.75.155111}. 
Hence $\tilde{r}_n$  is insensitive to systematic errors that arise while performing an unfolding of 
the spacing distribution, due to limited statistics and in small volumes. We have previously advocated 
its use for the first time to study localization properties of the eigenstates of the QCD Dirac 
operator~\cite{Pandey:2024goi}. Our results for $\langle \tilde{r} \rangle$ in bins 
of $\lambda/T$ obtained after averaging over gauge configurations at $T\gtrsim T_{pc}$ are 
summarized in Fig.~\ref{fig:rtildavslambda}.

We first focus on $T=164$ MeV which is just above the chiral pseudo-critical temperature. We observe 
that the $\langle \tilde{r} \rangle$ for eigenmodes in the lowest bin is in-between the values known for 
an uncorrelated and a Gaussian Unitary ensemble. We label them as \emph{intermediate} eigenmodes. 
For all other bins, the average value of $\langle \tilde{r} \rangle=0.602$~\cite{Pandey:2024goi} is 
consistent with expectations from a GUE~\cite{atas2013}. This categorization of Dirac eigenmodes 
into \emph{intermediate} and bulk can be easily done from Fig.\ref{fig:rtildavslambda} for other 
temperatures as well. At $T=164,~177$ and $195$ MeV, all eigenmodes which are below 
$\lambda/T=0.12,~0.33$ and $0.6$ respectively are designated as \emph{intermediate} 
eigenmodes, whereas the ones above these values are identified as bulk eigenmodes. 
Note that at these temperatures, the singlet $U_A(1)$ subgroup of chiral symmetry  
remains significantly broken~\cite{Gavai:2024mcj} instead only a remnant $Z_2$ subgroup 
is left intact.

Based on this information, we provide an interpretation for $\langle \tilde{r} \rangle$ of the 
\emph{intermediate} eigenmodes in terms of universal properties of random matrices. We refer to a 
recent work on understanding universal level spacing ratios in random matrices comprising of several 
independent blocks of known Wigner surmises~\cite{PhysRevX.12.011006}. 
Following their construction, we model the average spacing ratio for these \emph{intermediate} 
eigenvalues as a weighted sum denoted as $\langle \tilde{r} \rangle= w \langle \tilde{r} \rangle_1+ 
(1-w)\langle \tilde{r} \rangle_2 $, where $\langle \tilde{r} \rangle_{1,2}$ are the average spacing 
ratios from a single GUE block and two GUE blocks, respectively. Here, a single GUE block represents the 
contribution of bulk modes only, whereas a random matrix construction with two GUE blocks represents 
the scenario when the infrared eigenmodes that contribute to the chiral condensate and $U_A(1)$ 
breaking have admixture with the bulk eigenmodes. We remind here that the bulk eigenmodes survive 
even at temperatures when different subgroups of chiral symmetry are \emph{effectively} restored. 
Using the values of $\langle \tilde{r} \rangle=0.55(2)$ for the \emph{intermediate} eigenmodes at 
$T=164$ MeV and $\langle \tilde{r} \rangle_1=0.602$ and $\langle \tilde{r} \rangle_2=0.3992$
from Ref.~\cite{PhysRevX.12.011006}, we obtain the relative weight of the bulk eigenmodes to be $w=0.74(10)$. 
In contrast, the value for $\langle \tilde{r} \rangle$ for the lowest bin at $T=149$ MeV~\cite{Pandey:2024goi}
is consistent with one single GUE block because the $SU_A(2)$ and $U_A(1)$ subgroups of the chiral symmetry are 
explicitly broken and there are no remnant unbroken symmetries that can support the block-like decomposition. 

\subsection{Level spacing ratios in the region $T_{pc} < T < T_{U(1)}$ }
\label{sec:r_tilda_smallT}

We would now like to address the question of what happens to the level spacing ratios of the
\emph{intermediate} eigenmodes in the temperature regime when the $U_A(1)$ subgroup of chiral 
symmetry is broken whereas the non-singlet subgroup is \emph{effectively} restored. In this 
regime, the infrared eigenfunctions exhibit very interesting structural properties. We have 
earlier shown that at $T \sim 181$ MeV the \emph{intermediate} eigenmodes have fractal-like 
features and calculated their fractal dimension $D_f$~\cite{Pandey:2024goi}. It was observed 
that the median value of $D_f\sim 2.5$, which can be understood from the fact that these eigenmodes 
carry information of the remnant O(4) symmetry due to the \emph{effective} restoration of the 
non-singlet $SU_A(2)$ subgroup of the chiral symmetry.

The $U_A(1)$ subgroup should also get \emph{effectively} restored at a certain temperature 
$T_{U(1)}$, where the density of the smallest eigenvalues of the Dirac operator can be explained 
as arising due to a gauge background consisting of a dilute gas of instantons~\cite{Gavai:2024mcj}. 
In such a case, these eigenvalues should follow the level spacing distribution corresponding to an 
uncorrelated ensemble~\cite{Edwards:1999zm}. However as long as $U_A(1)$ is not \emph{effectively} 
restored, the infrared Dirac eigenmodes will have mixing with the remnant tail end of 
the bulk eigenspectrum, resulting in intermediate level statistics.  This is already evident 
in Fig.~\ref{fig:rtildavslambda} where the average value $\langle \tilde{r} \rangle$ of the 
eigenmodes in the deep infrared bin is decreasing as a function of increasing temperatures in 
the range $1.1 < T/ T_{pc} < 1.2$. At these temperatures, when the $U_A(1)$ is not yet effectively 
restored, we can model the lowest eigenvalues of the Dirac operator arising out of a matrix with 
a 2-block structure, one of them representing a matrix belonging to a GUE and the other 
to an uncorrelated ensemble. We have calculated the level spacing ratios in a 2-block system  
numerically and compared them with our lattice data, the details of which are discussed in 
Section~\ref{sec:DescribingMixedRatios}.

\subsection{Level spacing ratios in the region $T > T_{U(1)}$ }
\label{sec:r_tilda_largeT}
\begin{figure}[h]
    \centering
    \raisebox{-\height}{\includegraphics[width=0.48\textwidth]{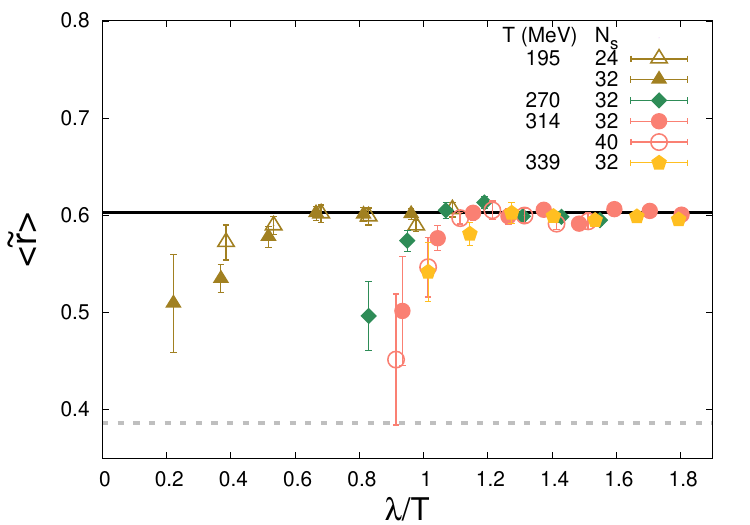}}
    \caption{ The variation of $\langle \tilde{r} \rangle$ in small bins in $\lambda/T$ at high temperatures 
    $T\gtrsim 270$ MeV when the $U_A(1)$ is \emph{effectively} restored, compared to the data at $195$ MeV. The 
    solid and dashed lines represent the expected $\langle \tilde{r} \rangle$ for the 
    eigenvalues with GUE and Poisson level statistics, respectively. For $T=195, 314$ MeV we show a 
    comparison between two different lattice volumes. }
    \label{fig:rtilda_high_T}
\end{figure}

At even higher temperatures $T > T_{U(1)}$, we expect that Dirac eigenvalues arising due to 
a gauge background consisting of a dilute gas of instantons will accumulate in the deep infrared 
part of the eigenvalue spectrum~\cite{Dick:2015twa, Ding:2020xlj}, thus separating from the bulk 
through a gap. We observe this gap opening up in the eigenvalue spectrum as the temperatures are 
increased beyond $195$ MeV.  Note that the values of $\langle \tilde{r}\rangle$ in the lowest 
bin at $\lambda/T=0.71, 0.80, 0.85$ for temperatures $T=270,314, 339 $ MeV respectively, 
have a large signal-to-noise ratio related to rare occurrences of such eigenvalues, hence not shown 
explicitly in Fig.~\ref{fig:rtilda_high_T}. We characterize all eigenmodes on an $N_s=32$ lattice, for 
which $\lambda/T <1.0,~1.1$ and $1.2$ at temperatures $T=270,~314$ and $339$ MeV respectively, 
as \emph{intermediate} eigenmodes since their $\langle \tilde{r} \rangle$ values are intermediate 
between a GUE and an uncorrelated matrix ensemble. Eigenvalues beyond these constitute the bulk 
eigenspectrum.

Extending the block-like construction discussed in the previous subsection, 
one can interpret the $\langle \tilde{r} \rangle$ values in the bins just above 
the gap as arising due to mixing between the chaotic and localized regions of the bulk 
eigenspectrum. This will be discussed in detail in section~\ref{sec:DescribingMixedRatios}. 
However, since the bulk spectrum is overwhelmingly chaotic with only 
a few localized eigenmodes in its tail, the $\langle \tilde{r} \rangle$ immediately 
approaches its GUE value within the first few bins as the temperature is increased. This is evidently 
different from the trend visible in the $\langle \tilde{r} \rangle$ data at $T = 195$ MeV, which is 
shown again in Fig.~\ref{fig:rtilda_high_T} for the sake of comparison. At this temperature, the gap 
is negligibly small, implying a larger fraction of the near-zero eigenmodes contributing to the 
$\langle \tilde{r} \rangle$, unlike the scenario at $T=270$ MeV. We have also calculated the 
same quantity at $T=195$ MeV for a smaller volume, results of which are compiled in 
Fig.~\ref{fig:rtilda_high_T}. A $50\%$ reduction in volume indeed leads to the depletion of  
\emph{intermediate} eigenvalues in the lowest bin, but the normalized level ratios are comparable 
in the other bins.  Anderson-like localization phenomenon for the Dirac spectrum~\cite{Holicki:2018sms, 
Kehr:2023wrs,Giordano:2021qav} has been studied in the same temperature regime discussed here. 
In order to discuss localization effects, one would need data on multiple different volumes, 
preferably larger than the ones we have at the highest temperatures that we study. The results for 
$\langle \tilde{r} \rangle$ on a larger $\sim (3.1~\rm{fm})^3$ volume at $T=314$ MeV is shown in 
Fig.~\ref{fig:rtilda_high_T}. We do observe volume dependence in the $\langle \tilde{r} \rangle$ 
for the most infrared eigenmodes, larger volumes bringing the value closer to that of an uncorrelated 
ensemble. However $\langle \tilde{r} \rangle$ values in the lowest few bins are still intermediate 
between a GUE and uncorrelated ensemble. We provide a more detailed interpretation of these 
\emph{intermediate} eigenmodes in section~\ref{sec:AndersonLoc} and discuss how well these can be 
compared to a 3D Anderson model near the mobility edge~\cite{PhysRevB.47.11487, PhysRevB.48.16979}.

\subsection{Interpreting the appearance of Dirac eigenvalues with intermediate level statistics}
\label{sec:DescribingMixedRatios}

\begin{figure}[h]
    \centering
    \raisebox{-\height}{\includegraphics[width=0.48\textwidth]{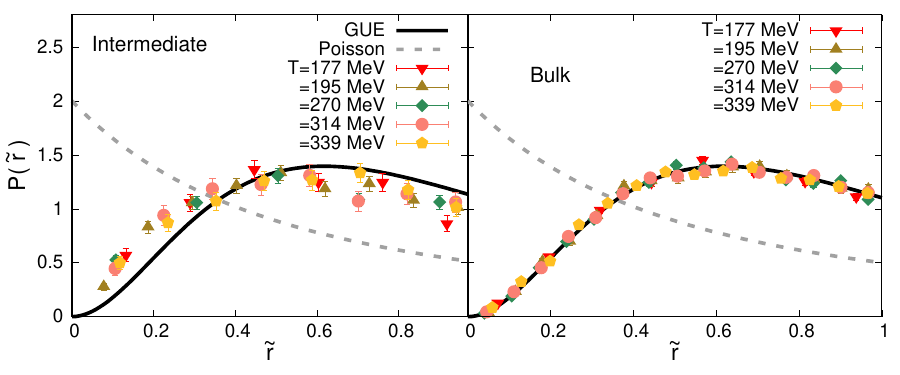}}
    \caption{Probability distribution of $\tilde{r}$ for the \emph{intermediate} (left) and bulk (right) eigenvalues at  different temperatures compared with its values for random matrices belonging to GUE (solid line) and 
    for uncorrelated eigenvalues following Poissonian level statistics (dotted line) respectively. }
    \label{fig:Prtilda_high_T_comp}
\end{figure}

In this section, we outline how the distribution of normalized level ratios $P(\tilde r)$ 
can give us important physical insights about the localization properties of the QCD Dirac spectrum. 
Before we proceed, we summarize our lattice results for $P(\tilde r)$ 
in different regions of the Dirac spectrum labeled as \emph{intermediate} and bulk, as a function of 
temperature, in the left and right panels of Fig.~\ref{fig:Prtilda_high_T_comp} respectively. The
$P(\tilde{r})$ for bulk eigenmodes for all the temperatures $T>T_{pc}$ are in excellent agreement with the 
prediction from random matrices belonging to GUE, as expected. The distribution $P(\tilde r)$ for the 
\emph{intermediate} eigenvalues is clearly different both from the expectations of uncorrelated eigenvalues 
as well as from a GUE. The questions we would like to address are, firstly, how to interpret these 
\emph{intermediate} level ratios in terms of a simple model of random matrices 
%whose eigenvalues have a non-trivial admixture with uncorrelated eigenvalues. 
and secondly, if a physically motivated model can explain the origin of these level ratios.

\begin{figure}[h]
    \centering
    \raisebox{-\height}{\includegraphics[width=0.48\textwidth]{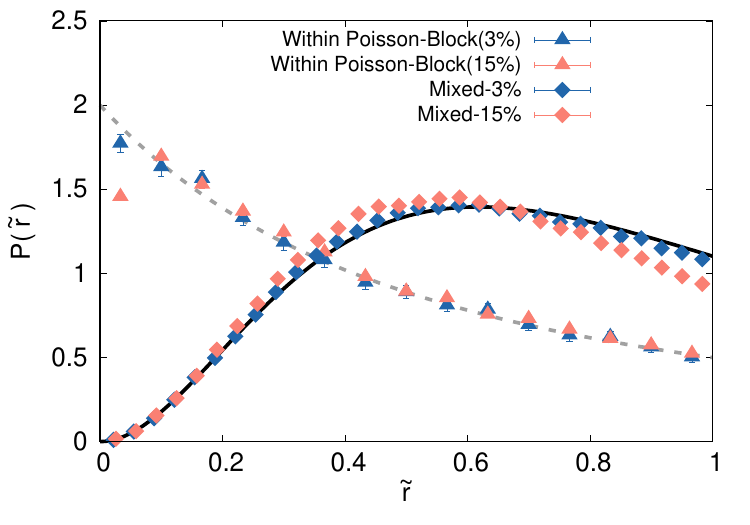}}
    \caption{Probability distribution of the normalized level ratios $P(\tilde{r})$ for our first matrix 
    model (see the text for the details). The $P(\tilde{r})$ for the newly added uncorrelated block and the entire 
    mixed matrix are shown as triangles and diamonds, respectively denoting $3\%$ (blue) and $15\%$ (orange) admixture. 
    The $P(\tilde{r})$ for eigenvalues belonging to uncorrelated and GUE matrices are shown as dashed and solid lines 
    , respectively.  }
    \label{fig:GUE-weight}
\end{figure}

We follow three different approaches to model the distribution of $\tilde r$ for the \emph{intermediate} 
eigenvalues in the same spirit as, but by extending concepts from the random matrix theory. We construct our 
models starting with a random matrix belonging to GUE with $0.3$ million eigenvalues. The probability 
distribution $P(\tilde{r})$ for the level ratios should follow the GUE prediction, which is confirmed and shown 
in Fig.~\ref{fig:GUE-weight} as a black solid line. As our first model, we construct a sufficiently small block of 
eigenvalues drawn from a uniform random distribution with a spacing distribution akin to uncorrelated eigenvalues 
following Poisson statistics whose size is $3\%$ and $15\%$ of the GUE block respectively. The eigenvalues of this 
block are chosen such that these lie within the range of the eigenvalues present in the GUE block except at the edges 
of this range, in order to allow for a complete mixture of the uncorrelated and GUE eigenvalues without any leakage. 
This is also done to mimic the QCD Dirac spectrum as closely as possible, where we do not observe any localized 
uncorrelated eigenvalues. A new eigenvalue $\lambda_{n'}$ is inserted between the existing eigenvalues 
$\lambda_{n}$ and $\lambda_{n+1}$ of the GUE block such that the spectrum of the new mixed ensemble is denoted 
by $\lambda_0, ~\lambda_1, ....,~\lambda_{n-1},~\lambda_{n},~\lambda_{n'},~\lambda_{n+1}, ~\lambda_{n+2},.....,
~\lambda_{N-1}$.  Here $\lambda_i, $ where $ i=0,..., N-1 $ are eigenvalues of our original GUE-block of size $N$. 
The three new level ratios that arise as a result are $r_{n_1} = \frac{\lambda_{n'}-\lambda_n}
{\lambda_n - \lambda_{n-1}}$ , ~$r_{n_2} = \frac{\lambda_{n+1}-\lambda_{n'}}{\lambda_{n'} - \lambda_{n}}$ and 
$r_{n_3} = \frac{\lambda_{n+2}-\lambda_{n+1}}{\lambda_{n+1} - \lambda_{n'}}$, which are normalized to give us
$\tilde{r}_{n_i} = \text{min}\left(r_{n_i},\frac{1}{r_{n_i}}\right)~,~i=1,2,3$. The new eigenvalues are accepted 
only if all three locally induced $\tilde{r}$ are compatible with the target distribution, i.e. $P_\text{GUE}
(\tilde{r})$, implemented through an independent accept–reject criterion. The acceptance probability of this newly 
selected eigenvalue is set to be $\prod_{i=1}^{3} \frac{P_\text{GUE}(\tilde{r}_{n_i})}{P^{\max}_\text{GUE}}$, where 
$P_\text{GUE}(\tilde{r})$ is the probability distribution of normalized ratios $\tilde{r}$ for matrices belonging 
to GUE whereas $P^{\max}_\text{GUE}$ is the maximum value possible for the distribution $P_\text{GUE}(\tilde{r})$.

Although the probability with which each new eigenvalue is accepted is the same as that of a GUE, for a sufficiently 
small block size of these newly chosen eigenvalues, it is obvious that the level repulsion between them is consistent 
with a Poisson distribution, as apriori no correlations exist between them. The probability distribution 
$P(\tilde{r})$ for the eigenvalues within this new block with block sizes $3\%$ and $15\%$ of the GUE-block 
respectively are also shown in Fig.~\ref{fig:GUE-weight}. From the plot, it is clearly evident that for a smaller 
block which is $3\%$ of the original GUE-block, the $P(\tilde r)$ agrees well with that known for a Poisson 
distribution. In the same figure, we have also shown $P(\tilde{r})$ for the full spectrum where existing 
eigenvalues from a GUE-block have an admixture of $3\%$ and $15\%$ of its size from the uncorrelated block, 
labeled as \emph{mixed}. From Fig.~\ref{fig:GUE-weight}, it is evident that this model cannot explain the 
$P(\tilde{r})$ for the \emph{intermediate} Dirac eigenmodes shown in Fig.~\ref{fig:Prtilda_high_T_comp} 
for smaller values of $\tilde{r}$.

\begin{figure}[h]
    \centering
    \raisebox{-\height}{\includegraphics[width=0.48\textwidth]{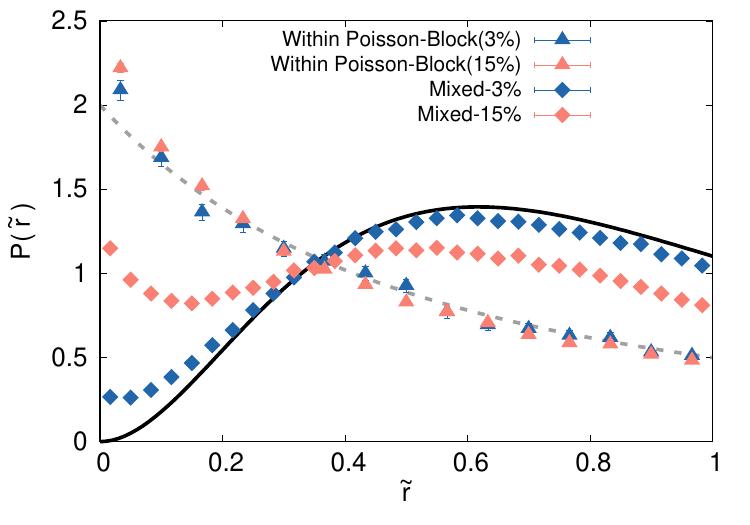}}
    \caption{ Probability distribution of the normalized level ratios $P(\tilde{r})$ for our second matrix 
    model (see the text for the details). The $P(\tilde{r})$ for the newly added uncorrelated block and the entire \emph{mixed} matrix are shown as triangles and diamonds, respectively denoting $3\%$ (blue) and $15\%$ (orange) admixture. The $P(\tilde{r})$ for eigenvalues belonging to uncorrelated and GUE matrices are shown as dashed and solid lines, respectively.  }
    \label{fig:Poisson-weight}
\end{figure}

Our second model is designed in a manner similar to the first one but in this case the randomly chosen new eigenvalues 
are accepted within the existing GUE-block with a probability $\prod_{i=1}^{3} \frac{P_\text{P}(\tilde{r}_{n_i})}{P^{\max}_\text{P}}$. Here $P_\text{P} (\tilde{r})$ is the probability distribution of normalized level ratios 
$\tilde{r}$ for uncorrelated eigenvalues following Poisson level statistics whereas $P^{\max}_\text{P}$ is 
the maximum possible value for the distribution $P_\text{P}(\tilde{r})$. In this approach, eigenvalues 
that are closely spaced will be more favored. We show the results of our calculation of $P(\tilde{r})$ 
in Fig.~\ref{fig:Poisson-weight}, when the newly added eigenvalue blocks are of sizes which are $3\%$ and $15\%$ 
respectively of the original GUE block. The $P(\tilde{r})$ of the newly added eigenvalue block is close to the 
value for uncorrelated eigenvalues, as expected. The $P(\tilde{r})$ of the entire matrix consisting of GUE and 
Poisson blocks is labeled as \emph{mixed} and is also shown in the same figure. For smaller values of 
$\tilde r$, the $P(\tilde{r})$ is intermediate between the values expected from Poisson and GUE level statistics 
respectively,  which is quite different from the trend observed for \emph{intermediate} eigenvalues of the QCD Dirac 
spectrum at all temperatures. This motivates us to formulate a third approach.

\begin{figure}[h]
    \centering
    \raisebox{-\height}{\includegraphics[width=0.48\textwidth]{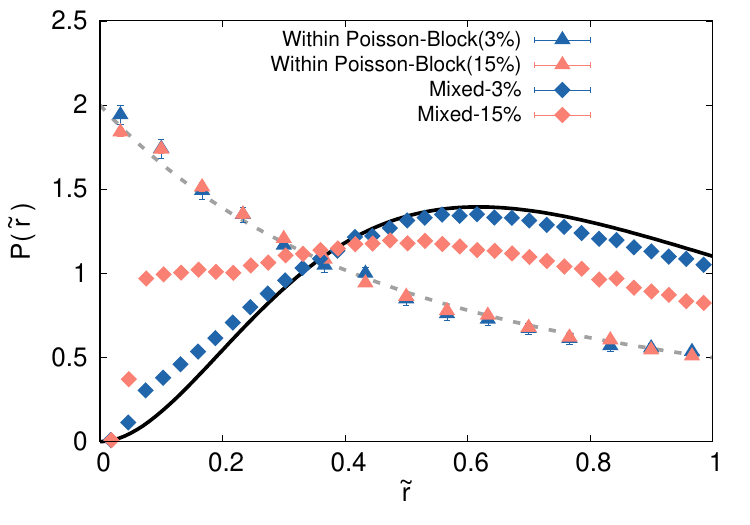}}
    \caption{ Probability distribution of the normalized level ratios $P(\tilde{r})$ for our third matrix 
    model (see the text for the details). The $P(\tilde{r})$ for the newly added uncorrelated block, and the entire 
    mixed matrix are shown as triangles and diamonds, respectively denoting $3\%$ (blue) and $15\%$ (orange) admixture. 
    The $P(\tilde{r})$ for eigenvalues belonging to uncorrelated and GUE matrices are shown as dashed and solid lines 
    , respectively. }
    \label{fig:Mixed-weight}
\end{figure}

\begin{figure}[h]
    \centering
    \raisebox{-\height}{\includegraphics[width=0.49\textwidth]{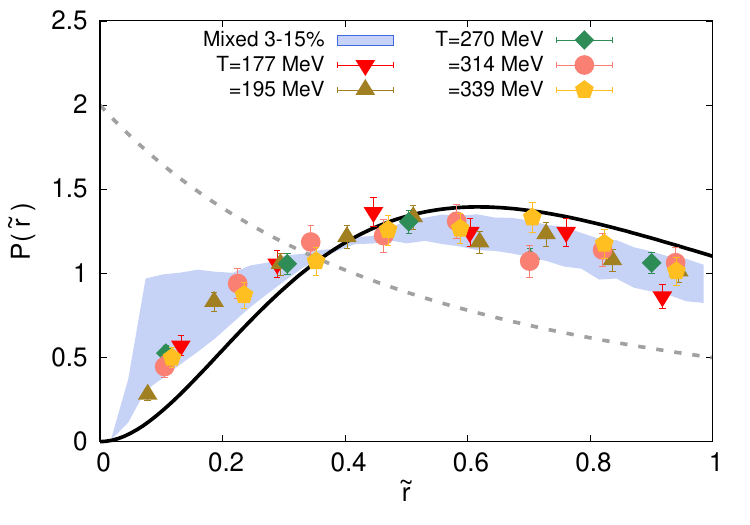}}
    \caption{The probability distribution of $\tilde{r}$ at different temperatures for \emph{intermediate} Dirac eigenmodes compared to the values obtained in our matrix model 3, shown as a blue band. The band represents values of $P(\tilde{r})$ corresponding to  $3$-$15 \%$ mixing described within our matrix model 3. The $P(\tilde{r})$ for eigenvalues belonging to uncorrelated and GUE matrices are shown as dashed and solid lines, respectively. }
    \label{fig:comparision-with-mixed-weight}
\end{figure}

Our third matrix model will focus on improving the modeling of $P(\tilde{r})$ to more closely mimic the lattice 
data at smaller values of $\tilde r$. Our lattice data for \emph{intermediate} eigenvalues shows some enhancement of 
$P(\tilde{r})$ compared to its magnitude in GUE for moderately small values of $\tilde r$, but the probabilities 
for the existence of eigenvalues with negligibly small spacing ratio $\tilde r \to 0$ are heavily disfavored. 
We model this fact in our random matrix construction by accepting newly generated eigenvalues with a probability 
$\prod_{i=1}^{3} P_\text{mix}(\tilde{r}_{n_i})$, where 
$P_\text{mix} (\tilde{r})\in   \{ \frac{P_\text{GUE}(\tilde{r})}{P^{\text{max}}_{\text{GUE}}}, \frac{P_{\text{P}}(\tilde{r})}{P^{\text{max}}_P}  \}$. Which of the two elements in this set decides the probability of accepting 
an eigenvalue depends on the values of $\tilde{r}_{n_1}, \tilde{r}_{n_2}, \tilde{r}_{n_3}$ and a threshold value 
of $\tilde{r}_\text{max}$ chosen in our study to be $ \tilde{r}_{\text{max}} = 0.05$. If the values of 
$\tilde{r}_{n_1},\tilde{r}_{n_2}, \tilde{r}_{n_3} < 0.05$ then the $P_{\text{mix}}(\tilde{r}_{n_i})= P_{\text{GUE}}
(\tilde{r}_{n_i})$ else $P_{\text{mix}}(\tilde{r}_{n_i})= P_{\text{P}}(\tilde{r}_{n_i})$. This selection criterion 
thus explicitly avoids selecting new eigenvalues which will lead to the proliferation of smaller values of $\tilde{r}$. 
The probability distributions $P(\tilde{r})$ for the newly added eigenvalues constituting a block of size $3\%$ and 
$15\%$ of the existing GUE block respectively for this \emph{mixed} model are shown in Fig.~\ref{fig:Mixed-weight}. 
For the smaller block size of the newly accepted eigenvalues, which comprises $3\%$ of the original GUE-block, the 
repulsion among the closely spaced eigenvalues is similar to expectations in a Poisson ensemble. In the same figure, 
we have also compiled our results for $P(\tilde r)$ of the entire matrix consisting of both GUE and uncorrelated blocks, which we consistently label as \emph{mixed}. The $P(\tilde r)$ for the eigenvalues chosen in this hybrid approach with a $3$-$15\%$ admixture can explain our lattice data for the normalized level ratios of \emph{intermediate} Dirac eigenvalues for a wide range of temperatures between $177$-$339$ MeV on our $N_s=32$ lattice. This is shown in Fig.~\ref{fig:comparision-with-mixed-weight} where the band represents the prediction from our model 3 with an admixture in the range $\sim 3$-$15\%$. Note, however, that the amount of admixture might vary within each small bin in $\lambda/T$ and its precise determination in the far infrared part of the spectrum will require further improvement in statistics.

We will now elaborate on a physical understanding of these models, i.e. why our model 3 can realistically 
describe the universal level spacing ratios for the \emph{intermediate} Dirac eigenvalues in different 
temperature regimes. Dyson provided an intuitive understanding~\cite{Dyson:1962brm} of the level spacing 
distribution of eigenvalues $\epsilon_i$ of an $N\times N$ random matrix by modeling them as fictitious Brownian 
point particles moving in one dimension at finite temperature, whose inverse is the Dyson index $\beta$. The potential 
that describes the Brownian motion of the eigenvalues consists of a confining as well as a repulsive part given 
by $V(\{\epsilon_i\})=\sum_{i=i}^{N} \epsilon_i^2-\sum_{i<j}\log \vert \epsilon_i-\epsilon_j \vert$, where 
$i,j$ label the eigenvalues. Performing a Monte-Carlo simulation of this system of $N=1000$ Brownian particles, 
the distribution of the normalized level spacing ratios of a random matrix belonging to GUE can be reproduced.

\begin{figure}[h]
    \centering
    \raisebox{-\height}{\includegraphics[width=0.48\textwidth]{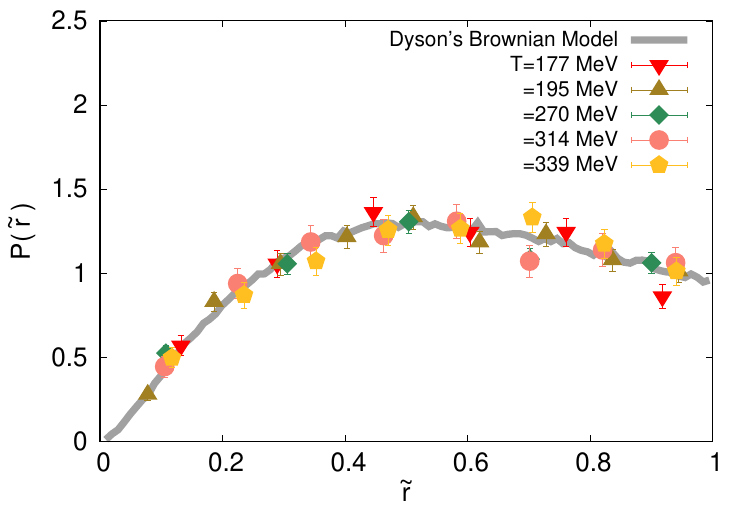}}
    \caption{The  probability distribution for $\tilde{r}$ obtained using Dyson's Brownian model for 
    eigenvalues $\{ \epsilon_i \}$ moving in the presence of a weak confining potential along with a log-repulsion between them, shown as a solid line. The probability distributions of $\tilde{r}$ for the Dirac eigenvalues with \emph{intermediate} level statistics, at different temperatures, are also shown (as points) for comparison. }
    \label{fig:dysonbrownian}
\end{figure}

Now, in order to understand the origin of \emph{intermediate} level statistics in terms of Dyson's 
construction, we replace the harmonic oscillator potential with another choice which causes a weaker 
confinement~\cite{Canali96, kravtsov1997new}.  The choice of different weak confining potentials was originally 
motivated in order to construct different classes of random matrix ensembles that have multi-fractal 
eigenfunctions and to understand what characteristic signatures of such multifractality are inherent in the 
spacing distribution~\cite{kravtsov1997new}. For one such choice of a confining potential, 
$V(\{\epsilon_i\})=A\sum_i \ln^2 \vert  \epsilon_i \vert$, we obtain a normalized level spacing ratio 
distribution which can very well explain our lattice data for the \emph{intermediate} Dirac eigenvalues, shown in 
Fig.~\ref{fig:dysonbrownian}. Though in our calculation we have chosen $A=0.2$, we have verified that 
level spacing ratio distributions are independent of the strength of the potential denoted by $A$, by 
varying it sufficiently such that $A\in [0,10]$. Another different choice of a weaker confining 
potential~\cite{Canali96}, $V(\{\epsilon_i\})=A \sum_i  \vert  \epsilon_i \vert ^ \alpha$ can explain our 
lattice data consistently as well.

In summary, by varying the fractional 
admixture of uncorrelated eigenvalues introduced within a large GUE matrix block according to a very 
specific criterion which is discussed above in model 3, the deviation of $\langle \tilde{r} \rangle$ 
from its GUE prediction can be very well understood for the infrared Dirac eigenvalues. The matrix model 
has a nice interpretation in terms of Dyson's construction of a Brownian model of eigenvalues as well. Our 
matrix model 3 is very general and versatile, which can describe the probability distribution $P(\tilde{r})$ for 
\emph{intermediate} eigenvalues appearing as a result of different physical mechanisms at work at different 
temperatures.

Similar analysis for the distribution of level ratios has been performed on a mixed ensemble of 
Poisson-GOE matrices in Ref.~\cite{PhysRevE.111.054213} by generalizing the techniques formulated in 
Ref.~\cite{PhysRevX.12.011006} for understanding mixing between eigenvalues of matrices belonging to 
different symmetry classes~\cite{RosenzweigPorter}. These works~\cite{PhysRevE.111.054213, 
PhysRevX.12.011006} derive the joint distribution of consecutive nearest-neighbor level spacings of 
a compound spectrum constructed out of spectra with different distributions. However for a 
Poisson-GUE system, there is no analytic solution for the normalized level spacing ratios in this 
approach, which have to be determined numerically. Our numerical approach is thus another new way 
to explain a mixed ensemble of matrices belonging to two different symmetry categories. We also note 
that modeling a mixed level-spacing distribution heuristically in terms of a best fit function with a 
few parameters~\cite{Brody:1981cx}, has been also discussed in the literature. A microscopic model consisting 
of a two-dimensional matrix mimicking a Poisson process weakly coupled with a random matrix belonging to 
different Wigner surmise~\cite{LenzHaake,KotaPGUE, Schierenberg:2012ut} has also been shown to fit a mixed 
distribution in terms of a single parameter that denotes the strength of the coupling.

\subsection{Understanding the origin of disorder and (de)localization in QCD Dirac eigenstates}
\label{sec:AndersonLoc}

Having interpreted the Dirac eigenmodes with intermediate level statistics as arising due to 
an admixture between eigenvalues with universal level spacing fluctuations of GUE and Poisson ensemble,  
we would now like to understand the microscopic origin of such a mixing.

We recall that in the Euclidean formalism, the partition function of QCD can be 
realized as a path integral with fermion fields satisfying anti-periodic boundary conditions along 
the compact imaginary time of extent $1/T$. As a result, the lowest eigenvalue of the Euclidean lattice 
Dirac operator for non-interacting quarks is $\sin (\pi/N_\tau)$. Turning on gauge interactions, 
the temporal part of the Dirac operator can be diagonalized at weak coupling. The diagonal elements at each lattice 
site turn out to be the lowest Matsubara mode shifted by the local Polyakov loop values~\cite{Schierenberg:2012ut}. 
A random matrix model of these local Matsubara modes with nearest-neighbor interaction closely resembles 
the tight-binding Anderson model~\cite{Schierenberg:2012ut}. At high temperatures, fluctuations of the 
Polyakov loop below its mean become rarer and occur at random spatial 
locations~\cite{Bruckmann:2011cc,Giordano:2015vla,Holicki:2018sms}, justifying its resemblance 
to a random disordered potential. However, the validity of these assumptions needs to be revisited 
for temperatures $T_{pc} < T < T_{U(1)}$. We will address this point in this section.

\begin{figure*}[t]
%\begin{center}
    \includegraphics[scale=0.45]{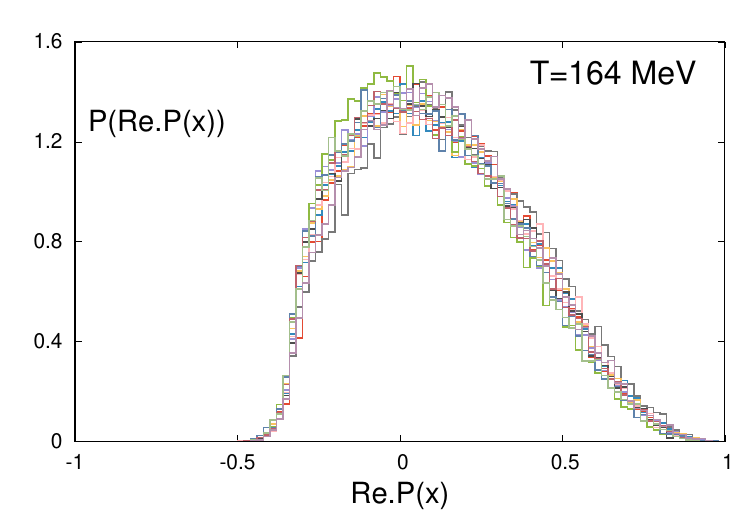}
    \includegraphics[scale=0.45]{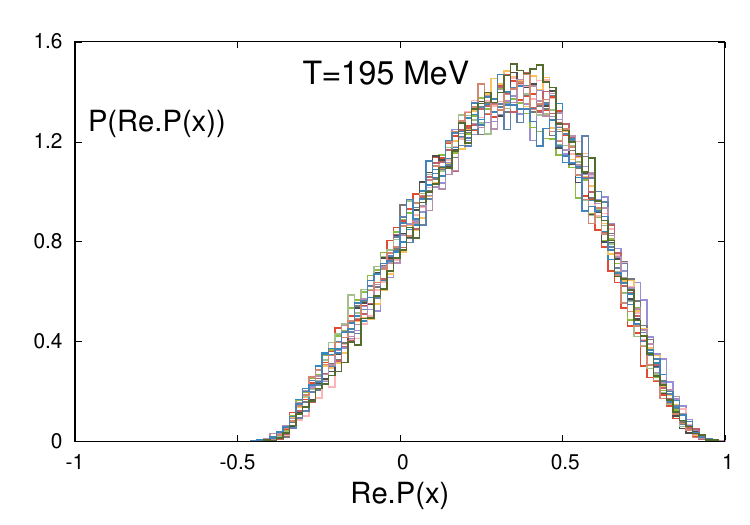}
    \includegraphics[scale=0.45]{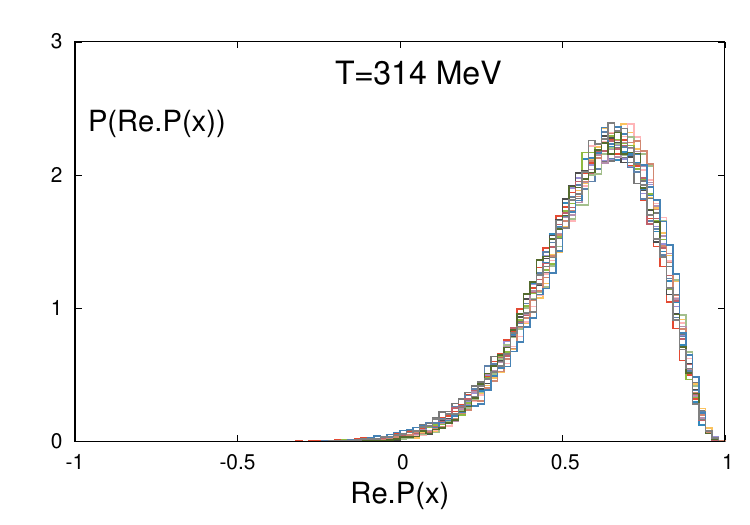}
    \caption{Probability distribution of the $\text{Re.}P(\textbf{x})$ for statistically independent 
    gauge configurations, calculated after performing Wilson flow renormalization at three 
    different temperatures above the chiral crossover transition. The three panels, from left to right, correspond to 
    temperatures $T=164, 195$ and $314$ MeV.}
    \label{fig:polyakov_prob}
%\end{center}
\end{figure*}

In the regime $T\gtrsim T_{pc}$, the regions with strong fluctuations in the local values of the Polyakov 
loop are typically (anti)correlated with the probability density for the occurrence of 
instanton-dyons~\cite{Larsen:2021ppf} which form a semi-classical gas~\cite{Larsen:2018crg,Larsen:2019sdi}. 
Hence, the origin of disorder at temperatures $T_{pc}$ is largely driven by fluctuations about the topological 
vacuum in QCD.  In this regime, the temperature dependence of topological susceptibility is different from that of a dilute 
instanton gas~\cite{Petreczky:2016vrs}. This is the same region where $U_A(1)$ remains badly broken precisely 
because of these non-trivial topological fluctuations. When temperatures are $T\gtrsim 1.5~T_{pc}$, the holonomy 
becomes trivial and the sites with discernible random fluctuations of the Polyakov loop values are no longer 
correlated with the fluctuations of the topological vacuum in QCD. Incidentally, this is the regime when $U_A(1)$ 
is believed to be \emph{effectively} restored and randomness in the fluctuations of the local Polyakov loop should be 
solely driven by the inherent chaotic nature of color interactions~\cite{Blaizot:2008nc, Guin:2025lpy}.

\begin{figure*}[t]
%\begin{center}
    \includegraphics[scale=0.392]{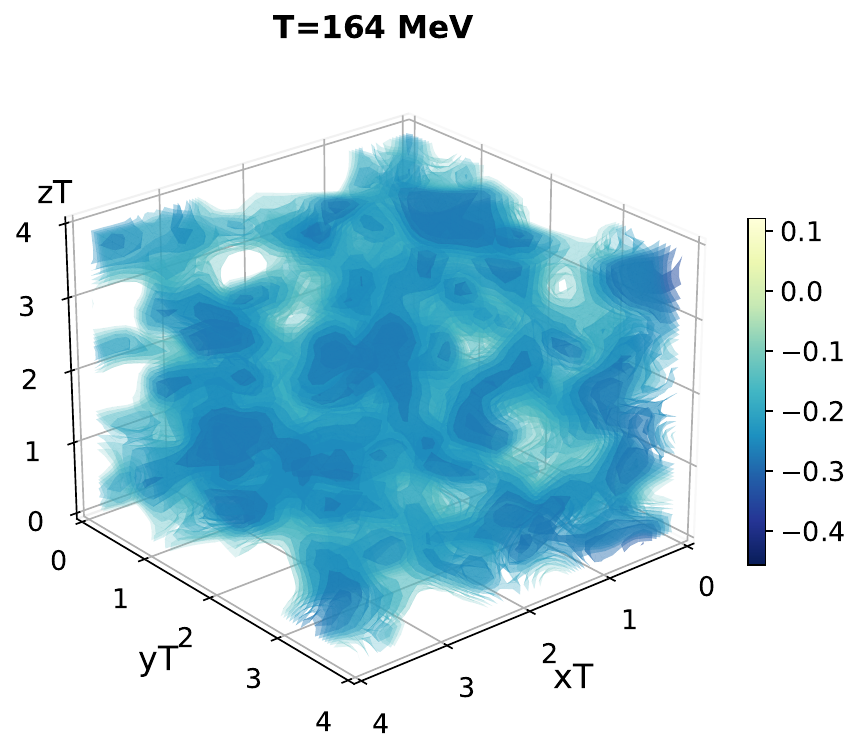}
    \includegraphics[scale=0.392]{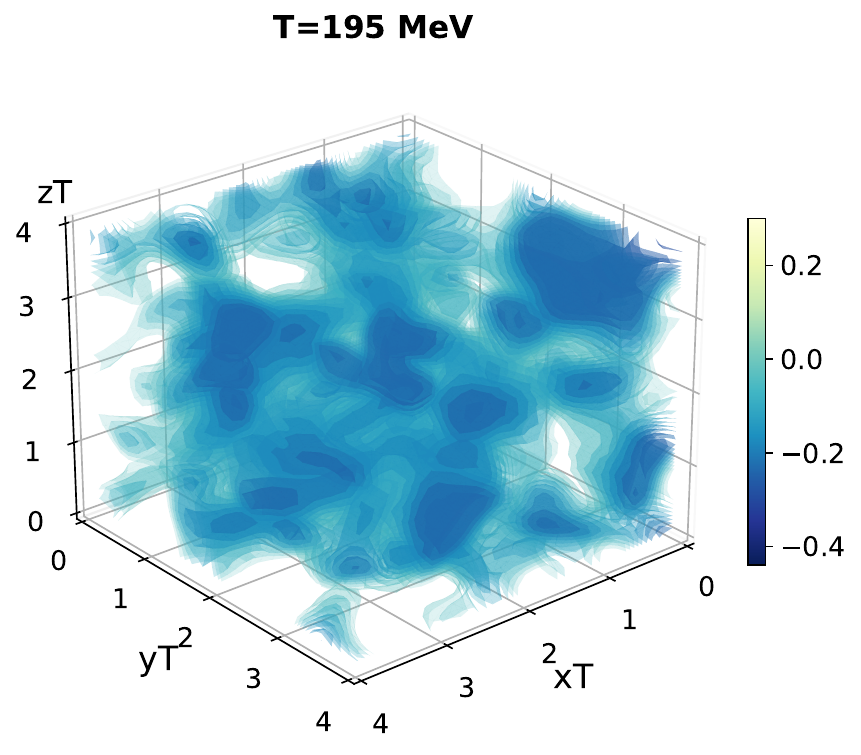}
    \includegraphics[scale=0.392]{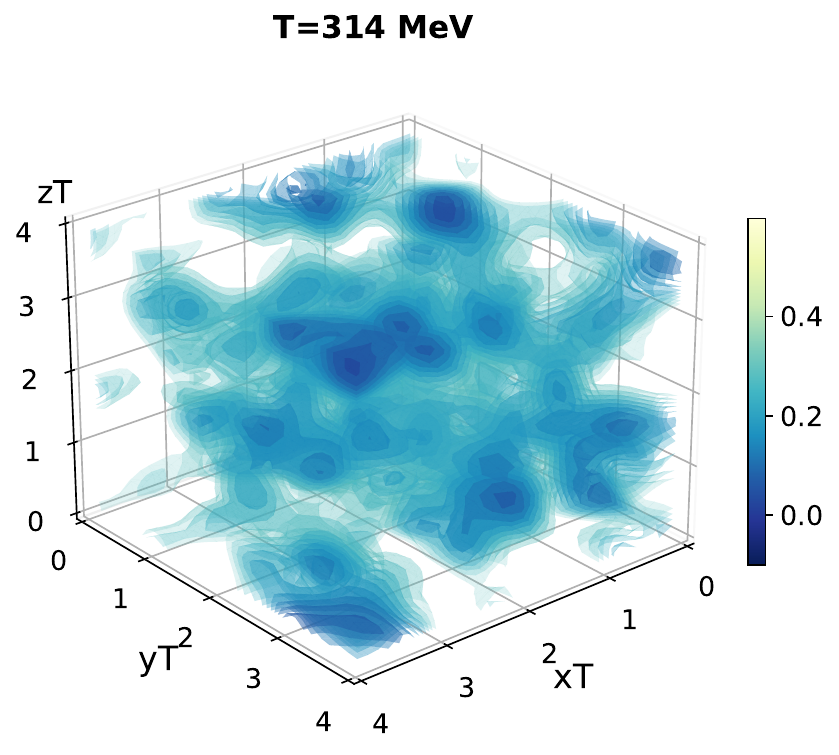}
    \caption{Visual representation of the local fluctuations in the real part of the Polyakov loop 
    $\text{Re.}P(\textbf{x})$ which are below its mean value, obtained after performing Wilson flow 
    renormalization on a typical gauge configuration. The lightest color in the palette represents its mean 
    value in that particular configuration. The three panels from left to right correspond to three different 
    temperatures $T=164, 195$ and $314$ MeV.} 
    \label{fig:vis}
%\end{center}
\end{figure*}

We also review the literature to clarify some important points that are not often discussed 
in the study of Anderson localization in QCD. Firstly, since the gauge field configurations represent 
a strongly interacting many-body system at $T_{pc} < T < T_{U(1)}$, it is not clear apriori whether the 
noise inherent in it can be considered as random, as required in the Anderson model. The generalization 
of Anderson localization in interacting systems, known as many-body localization (MBL), has been discussed 
as a finite temperature phase transition~\cite{BASKO20061126} and is currently under intensive 
review~\cite{Nandkishore:2014kca,RevModPhys.91.021001}. 
The characteristic signatures of MBL have been shown to be manifest in many single-particle observables as 
well~\cite{SoumyaBeraMBL2015}. It is important to note that the disorder inherent in the gauge fields in QCD 
as a function of temperature is not quenched (static) but dynamically generated due to self-interactions, 
similar to a scenario discussed in the context of interacting lattice spin models~\cite{MoessnerDisorderFreeLoc}.

Secondly, we would like to remind here that when we calculate eigenvectors of the Dirac 
operator in the presence of a noisy gauge background, we are solving for relativistic single-particle states.
This is different from a conventional Anderson model, which discusses non-relativistic electrons in 
the presence of a quenched random disorder. Anderson transition has been studied in three-dimensional 
non-interacting Dirac semimetals in the presence of quenched disorder~\cite{SDasSharma3DAL2015}. It was reported 
that such systems have three distinct ground states, i.e., an incompressible semimetal, a compressible diffusive 
metal, and an insulator, which is in contrast to a conventional Anderson model that describes a transition between 
only the latter two phases. In presence of a small amount of disorder, there is a quantum phase transition from a 
semimetal to a dirty metal phase, which, on increasing the disorder strength by about six times, goes over to an 
insulating phase through the conventional Anderson semimetal-insulator transition. In our earlier study in 
quenched SU(3) gauge theory at $T\sim 600$ MeV, we did not observe any localized Dirac eigenstate 
in the deep infrared part of the spectrum~\cite{Pandey:2024goi}, which characterizes an insulating phase. 
This absence of localized eigenmodes may be attributed to the weakening of the disorder in the gauge fields. 
Hence, it is imperative that we make a more quantitative estimate of the amount of disorder inherent in the 
gauge fields in QCD, as a function of temperature.

For this purpose, we have calculated the renormalized Polyakov loop values at each spatial site of the lattice 
after performing gradient flow, details of which are mentioned in Sec.~\ref{sec:Num}. We show the probability 
distribution of renormalized values of the real part of the Polyakov loop at each site $\mathbf{x}$, denoted as 
$\text{Re.}P(\textbf{x})$ for a large set of gauge configurations, superposed together, and at different 
temperatures in Fig. \ref{fig:polyakov_prob}. We observe that the width of these distributions is decreasing 
with increasing temperatures, demonstrating that gauge fields become smoother at higher temperatures. 
At higher temperatures $T=314$ MeV, the large negative fluctuations in $\text{Re.}P(\textbf{x})$ about its 
average value become rarer, resulting in a clear separation among these potential wells which are located 
at random sites. In contrast, close to the crossover transition at $T=164$ MeV, the fluctuations in the 
renormalized values of the Polyakov loop are much more frequent and correlated with each other, 
which is qualitatively manifest in the density plot in Fig.~\ref{fig:vis}. At an intermediate temperature 
$T= 195$ MeV, the confining wells still have significant overlap with each other which is also evident from Fig.~\ref{fig:vis}. 

\begin{figure}[h]
    \centering
    \raisebox{-\height}{\includegraphics[width=0.48\textwidth]{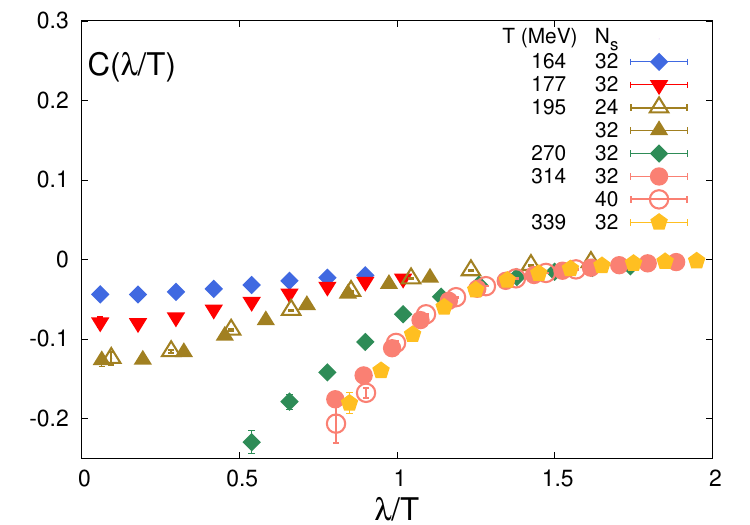}}
    \caption{The correlation $C(\lambda/T)$ between the local density of the Dirac eigenstates and the local fluctuation of $\text{Re.}P(\textbf{x})$, evaluated in small bins of $\lambda/T $ at different temperatures. For $T=195, 314$ MeV, we show a comparison between two different lattice volumes. }
    \label{fig:corrilation}
\end{figure}

In order to quantify how well these deep negative fluctuations of the renormalized Polyakov loop $\text{Re.}
P(\textbf{x})$ about its mean are correlated with the Dirac eigenstates at each temperature, we calculate the 
quantity $C_n = \sum_{\textbf{x}} |\Psi_{n}(\textbf{x})|^2 \big[\text{Re.}P(\textbf{x}) - \langle \text{Re.}
P(\textbf{x})\rangle \big]$.  Here the norm 
$|\Psi_n(\textbf{x})|^2 = \sum_{\tau=0}^{N_\tau -1}  |\psi_n(\textbf{x},\tau)|^2$ 
at each spatial site $\mathbf{x}$ on the lattice is obtained by summing over the density 
of the $n^{th}$ eigenstate of the overlap Dirac operator in four dimensions along the temporal direction 
$\tau$.  We calculate the net correlation $C(\lambda/T)$ by performing a binning of the $C_n$, 
calculated for each eigenstate labeled by $n$, in small bins in $\lambda/T$ and performing an average 
over the gauge configurations for different temperatures, the results of which are compiled in 
Fig.~\ref{fig:corrilation}. It is evident from the figure that eigenvalues in the 
\emph{intermediate} region of the Dirac spectrum are more strongly correlated with the wells 
arising due to fluctuating renormalized Polyakov loop values as compared to the bulk 
eigenmodes. Moreover at $T\leq 195$ MeV, the correlation between \emph{intermediate} 
Dirac eigenstates with the local Polyakov loop fluctuations is weaker compared to temperatures 
$T>250$ MeV. This is due to the fact that these well-like fluctuations at different lattice sites 
are no longer rare uncorrelated events at $T\leq 195$ MeV, as evident from Fig.~\ref{fig:vis}. 
As a result, \emph{intermediate} eigenstates are delocalized 
over several such locations. This is also evident from the spectral properties~\cite{Alexandru:2024tel} 
and fractal-like feature~\cite{Pandey:2024goi} of these \emph{intermediate} eigenstates at $T<T_{U(1)}$ 
observed in previous works. In conclusion, the localization of the \emph{intermediate} Dirac eigenstates 
in wells created due to fluctuations of $\text{Re.}P(\textbf{x}) $ is only apparent when $T>T_{U(1)}$.

We next would like to understand the sensitivity of this correlation measure to the change 
in volume. Whereas at $T=195$ MeV we observe very little volume dependence in $C(\lambda/T)$,  
this correlation measure increases with increasing volume for the \emph{intermediate} 
eigenmodes at $T=314$ MeV. This reinforces our earlier conclusion. Nonetheless, it will 
be important to perform a volume extrapolation of this quantity at the higher temperatures 
for the \emph{intermediate} eigenstates. The bulk eigenstates, on the other hand, are delocalized 
chaotic states and hence do not feel these tiny local gauge fluctuations, resulting in no volume dependence.

%This demonstrates that indeed random local fluctuations in $\text{Re.}P(\textbf{x})$ act as confining potentials for these low eigenmodes at higher temperatures.

\subsection{Structural features of the Dirac eigenfunctions from Thouless conductance} 
\label{sec:conductance}

Apart from the local correlations that are observed, e.g., in the spacing ratios, it is 
also important to identify global correlations that exist in the eigenspectrum. This will 
allow us to identify the physical origin of correlations that result in distinct structural 
changes in the eigenfunctions of the Dirac operator as a function of temperature.

\begin{figure}[h]
    \centering
    \raisebox{-\height}{\includegraphics[width=0.48\textwidth]{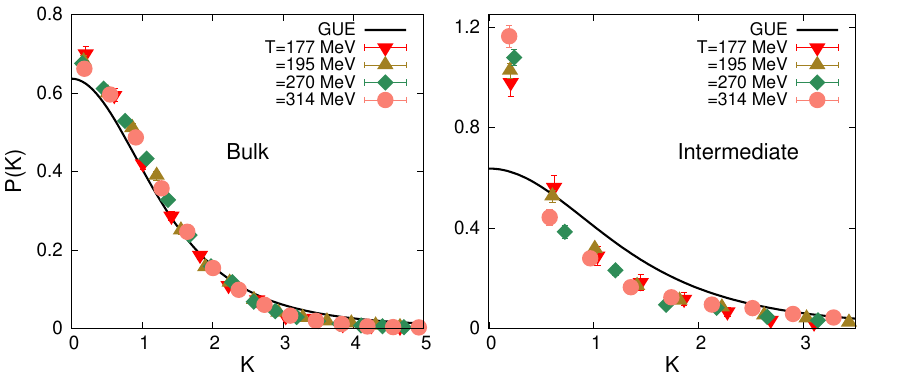}}
    \caption{The curvature distribution of the bulk (left) and \emph{intermediate} (right) Dirac eigenmodes as a 
    function of temperature. The black solid line represents the expected probability distribution for the curvature $K$ for eigenvalues of random matrices belonging to a GUE.}
    \label{fig:curvatureDistr}
\end{figure}

One such quantity sensitive to global correlations is the Thouless conductance, 
which is measured in terms of the average shift in eigenvalues of the Hamiltonian 
of a quantum system due to a change in the boundary conditions along a spatial 
direction~\cite{JTEdwards_1972}.  According to the Thouless criterion, one could then 
categorize eigenstates as extended or localized if the average shift is much larger 
than the average level spacing or vice versa. In our context, if the fermion fields 
$\psi_i$ satisfy the boundary condition $\psi_i(x+L)=\psi_i(x) \rm{e}^{i\eta}$ along 
any of the spatial directions of length $L$, this would modify the Dirac operator. 
As a result, its $i$-th eigenvalue should acquire a curvature $K_i$ defined as 
$K_i=\Big\vert \frac{\partial ^2 \lambda_i}{\partial \eta^2} \Big\vert$. The twist $\eta$ 
should be infinitesimally small, i.e. $\eta\to 0$. We have chosen $\eta$ such that the 
change in eigenvalues after applying the twist is larger than the numerical precision of 
$\sim 10^{-10}$ with which the eigenvalues are calculated. We have chosen $\eta = 0.005$ 
for our calculations. We have also verified that using $\eta = 0.01$ does not lead to any 
change in our results. We then compute the probability distribution of the curvature values, 
$P(K)$, separately for the \emph{intermediate} and bulk eigenmodes. This is achieved by constructing 
appropriate bins for the normalized curvature values, $K_i \rightarrow K_i/\langle K \rangle$, 
where $\langle K \rangle$ denotes the average curvature for the \emph{intermediate} and bulk modes, 
respectively. After performing an ensemble average over available gauge configurations, the 
resulting distributions $P(K)$ for four different temperatures on $N_s=32$ lattice are shown in 
Fig.~\ref{fig:curvatureDistr}.

Bulk eigenstates, which are delocalized over the entire volume, should acquire larger curvature 
upon applying a twist compared to \emph{intermediate} eigenstates whose inverse participation ratios do 
not scale inversely with volume. There can, however, be rare large fluctuations in the curvature distribution 
due to localized eigenstates that might be situated close to the boundaries of the lattice box. Our lattice data 
for the curvature distribution of the \emph{intermediate} eigenmodes, shown in the right panel of 
Fig.~\ref{fig:curvatureDistr}, are consistent with these expectations. We have also compared 
the curvature distribution of bulk eigenmodes with the expectation from a GUE~\cite{vonOppen:1994zz}, 
given as $P(K)=\frac{2}{\pi}\frac{1}{\left(K^2+1\right)^2}$. As evident from Fig.~\ref{fig:curvatureDistr}, 
there is a fairly good agreement between $P(K)$ for bulk eigenmodes and the prediction from a GUE, for all 
temperatures we have studied.

\begin{figure}[h]
    \centering
    \raisebox{-\height}{\includegraphics[width=0.48\textwidth]{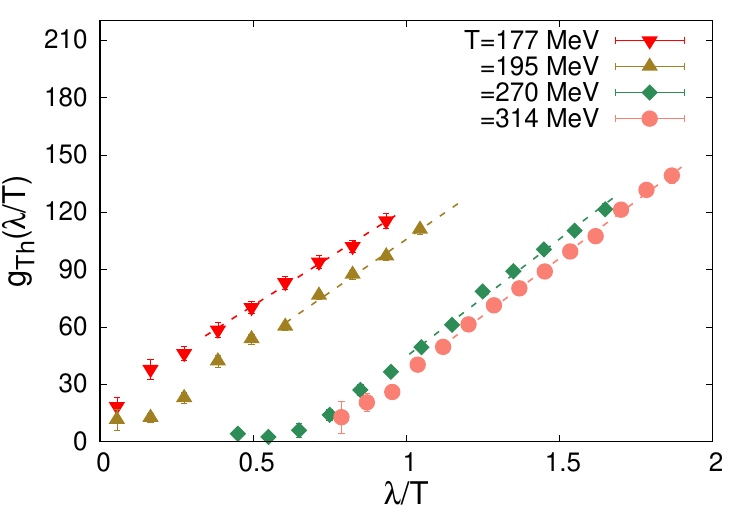}}
    \caption{The Thouless conductance for different eigenvalue bins calculated at temperatures $T=177, 195, 270, 314$ MeV respectively. The dashed line represents a linear fit to the $g_\text{Th}(\lambda/T)$ for bulk eigenmodes.}
    \label{fig:thouless}
\end{figure}

We next calculate the Thouless conductance, 
defined as $g_\text{Th} = \frac{1}{\left\langle s \right\rangle} \langle 
 K_i \rangle_{\eta \rightarrow 0}$ in terms of the curvature $K_i$ acquired by 
the $i$-th eigenstate subject to an infinitesimally small twist $\eta$. 
Calculating averages for the curvatures $\langle K_i \rangle$ and level spacings 
$\langle s \rangle$ in small bins in $\lambda/T$ at four different temperatures, 
on our $N_s=32$ lattice we summarize our results for $g_\text{Th}$ in 
Fig.~\ref{fig:thouless}. The $g_\text{Th}$ is non-vanishing in the \emph{intermediate} 
region, which implies that these eigenstates are not exponentially localized. Performing 
a fit to the data for bulk eigenvalues with an ansatz function that is linear in $\lambda$,  
we obtain a positive slope $d g_\text{Th}(\lambda/T)/d (\lambda/T)$ which has mild sensitivity 
to the change in temperature. The functional dependence of the Thouless conductance on $\lambda/T$, 
in the bins containing \emph{intermediate} eigenvalues, is different compared to those with bulk 
eigenmodes. Thus $g_\text{Th}$ allows us to unambiguously disentangle the bulk eigenvalues 
from the rest.

\begin{figure}[h]
    \centering
    \raisebox{-\height}{\includegraphics[width=0.48\textwidth]{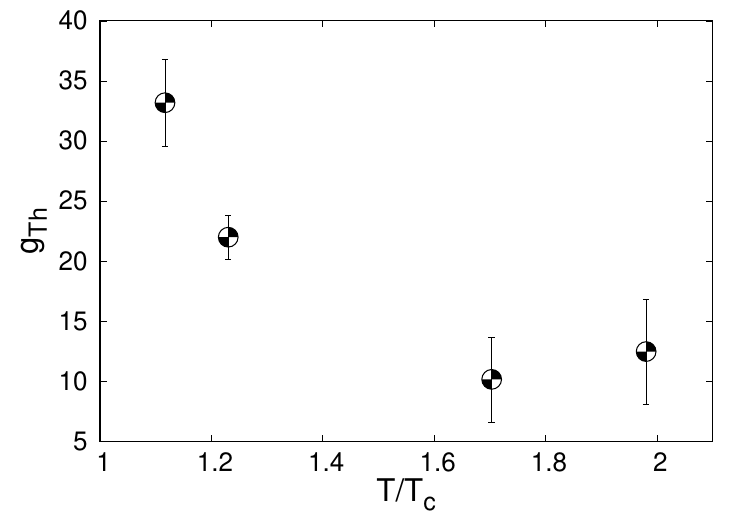}}
    \caption{Thouless conductance for the \emph{intermediate} eigenmodes which have similar values of $C(\lambda/T)$, shown as a function of temperature.}
    \label{fig:thouless_lower_bin}
\end{figure}

We show our results for $g_\text{Th}$ as a function of temperature, only due to \emph{intermediate} eigenmodes 
in Fig.~\ref{fig:thouless_lower_bin}. We have chosen only the lowest few \emph{intermediate} eigenvalues, which are 
strongly correlated with the gauge field fluctuations, quantified in terms of $C(\lambda/T)$. It is clear that with 
increasing temperatures, these \emph{intermediate} eigenmodes get more and more localized within the wells formed 
due to fluctuations in $\text{Re.}P(\textbf{x})$, leading to a reduction in the Thouless conductance. Value of 
$g_\text{Th}$ drops very fast between $1.1$-$1.5~T_{pc}$, saturating at higher temperatures $T>T_{U(1)}$, albeit with 
large errors due to fewer eigenmodes in the lowest bins. It is important to note that the \emph{intermediate} 
eigenmodes at the highest temperature get further localized within the wells due to local fluctuations of the renormalized 
Polyakov loop with increasing volumes. One would thus expect the $g_\text{Th}$ to show very little sensitivity to 
volume. However, at least three different lattice volumes would be required for quantifying the volume scaling of 
$g_\text{Th}$. At present, the statistics is not sufficient to perform a rigorous finite volume analysis of this 
observable, which will be addressed in a future work.  Nonetheless $g_\text{Th}$ mimics as an \emph{order parameter} 
for the \emph{effective restoration} of $U_A(1)$ and the point of inflection of the curve that characterizes 
$g_\text{Th}$ versus temperature could give us an estimate of $T_{U(1)}$. Noting that $U_A(1)$ is not an exact 
symmetry, one cannot construct an order parameter that conventionally characterizes a thermal phase transition within 
the Landau paradigm. We thus propose that the \emph{effective} restoration $U_A(1)$ should be inferred from the change 
in the structural properties of the Dirac eigenstates, which can be quantified in terms of observables like 
$g_\text{Th}$.

\section{Summary and Outlook}

In this work, we have performed a detailed lattice study of the properties of infrared eigenstates of the 
massless Dirac operator on thermal gauge ensembles, in order to understand the high-temperature 
phase of a strongly correlated system described by QCD. We particularly focused on the question: how can we identify 
the \emph{effective} restoration of the anomalous U(1) subgroup of chiral symmetry using spectral and structural 
properties of the Dirac eigenstates, and what are its physical consequences? We have performed our calculations with 
an extensive set of gauge configurations generated using M\"{o}bius domain wall fermion discretization, over a 
range of temperatures up to $\sim 2~T_{pc}$. Using normalized level spacing ratios of the eigenvalues of the massless 
1-particle states realized using the overlap Dirac operator, we have unambiguously identified those eigenmodes 
which have intermediate level statistics from the overwhelming majority of the bulk eigenstates. These 
bulk eigenstates have level repulsions consistent with random matrices belonging to a GUE. Though we 
have earlier provided an explanation for the existence of these \emph{intermediate} eigenstates just above 
$T_{pc}$ as arising due to the restoration of the non-singlet part of the almost exact chiral symmetry for QCD 
with two light quark flavors~\cite{Pandey:2024goi}; however, we show here that their persistence at very high 
temperature has a different physical origin. 

By studying the correlation of these \emph{intermediate} eigenstates with fluctuations of the 
real values of the renormalized Polyakov loop about its mean, we could clearly show their distinct physical 
properties at high temperatures $T> 250$ MeV, where the axial U(1) symmetry is expected to be effectively 
restored, compared to $T< 200$ MeV, where it remains broken.  In fact, the density plots for the real part of the 
renormalized Polyakov loop reveal that at $T> 250$ MeV, there are deep localized fluctuations at random sites of the 
lattice which are uncorrelated with each other, mimicking random disorder as discussed in the context of the Anderson 
model of electrons. We will study the finite volume effects more carefully in the future to understand whether these 
infrared eigenstates with intermediate level ratios describe the physics near the mobility edge in an 
Anderson-like transition. We also showed that fluctuations in the real values of the renormalized 
Polyakov loop at different lattice sites are strongly correlated at $T<200 $ MeV, quite unlike the original 
proposal of a random disorder by Anderson. Our study hints at the fact that localization of the most infrared Dirac 
eigenstates will start to become more prominent when the axial U(1) subgroup of chiral symmetry is effectively 
restored. 

We then provide an explanation for the existence of these Dirac eigenstates with intermediate level 
statistics within a matrix model. By introducing additional uncorrelated eigenvalues within a large random matrix 
belonging to a GUE with a very specific accept/reject criterion, we showed that normalized level ratios of this 
enlarged matrix can beautifully explain intermediate level ratios of the QCD Dirac eigenvalues shown in 
Fig.~\ref{fig:comparision-with-mixed-weight}. We also discussed an intuitive understanding of such a matrix model 
construction in terms of a Brownian model describing the dynamics of the eigenvalues. Unlike the eigenvalues of a 
random matrix belonging to a GUE whose level separations arise due to its Brownian motion in a thermal bath with a 
weak log-repulsion and a strong confining harmonic potential, the intermediate eigen level ratios can be explained 
if the confining potential is weakened.

We also, for the first time, calculate the analogue of the Thouless conductance $g_\text{Th}$ of the Dirac 
eigenstates which quantifies their structural rigidity when subject to a twist applied along one of the spatial 
boundaries of the lattice box. We use the curvature distribution and the $g_\text{Th}$ to also independently 
distinguish between the \emph{intermediate} eigenstates and the bulk.  Calculating the $g_\text{Th}$ only for the 
\emph{intermediate} eigenstates that are strongly correlated with the local Polyakov loop fluctuations, 
we demonstrate its gradual decrease as a function of temperature. The point of inflection of this curve 
could give us an estimate of the temperature where $U_A(1)$ is \emph{effectively} restored, and we will 
study this observable in the thermodynamic limit to firmly establish our claim in a future work.

\section*{ACKNOWLEDGMENTS}

We are grateful to Robin Kehr and Lorenz von Smekal for discussions on related topics during the course of 
this work. We acknowledge support from the Institute of Mathematical Sciences, and the computing time allocation 
at the Institute cluster. This research was supported in part by the International Centre for Theoretical Sciences 
(ICTS) for participating in the program $-$ Generalised symmetries and anomalies in quantum phases of matter 2026 
(ICTS/GSYQM2026/01). Our GPU code is in part based on some of the publicly available QUDA~\cite{Clark:2009wm} and 
Grid Python Toolkit~\cite{GPT} libraries.

\bibliography{Paper_RHS.bib}

\end{document}